\documentclass[10pt]{article}
\usepackage{times}
\usepackage{amsmath}
\usepackage{amssymb}
\usepackage{stmaryrd}
\usepackage[english]{babel}
\usepackage{supertabular }

\newcommand{\syntaxdef}{::=}
\newcommand{\procdef}{\displaystyle \mathop{=}^{\mbox{\scriptsize def}}}
\newcommand{\nattype}{\mbox{Nat}}
\newcommand{\qubittype}{\mbox{Qubit}}
\newcommand{\qvar}[1]{#1:\qubittype}
\newcommand{\cvar}[1]{#1:\nattype}
\newcommand{\declvar}[2]{#1: #2}
\newcommand{\findeclvar}{\pmb\ .\ }
\newcommand{\debutbloc}{\pmb [\ }
\newcommand{\finblocse}{\pmb ]}
\newcommand{\finbloc}{\ \pmb ]}
\newcommand{\prefix}{.}
\newcommand{\seq}{\ ;}
\newcommand{\para}{\parallel}
\newcommand{\nondet}{+}
\newcommand{\cond}[2]{#1 \rightarrow #2}
\newcommand{\restrict}[1]{\backslash#1\ }
\newcommand{\restrictg}[1]{\backslash\{#1\}\ }
\newcommand{\envoi}[1]{\ !#1\  }
\newcommand{\recep}[1]{\ ?#1\  }
\newcommand{\term}{\mbox{\it end}}
\newcommand{\stopproc}{\mbox{\it nil}}
\newcommand{\mathoparrow}{\mathop{\longrightarrow}}
\newcommand{\transition}{\hspace{10pt}\longrightarrow\hspace{10pt}}
\newcommand{\transitionse}{\longrightarrow} 
\newcommand{\atransition}[1]
{\hspace{10pt}\underrightarrow{\hspace{10pt}#1\hspace{10pt}}\hspace{10pt}}

\newcommand{\petiteatransition}[1]{\mbox{$\displaystyle\ \mathoparrow^{#1}\ $}}
\newcommand{\tautransition}
{\hspace{10pt}\underrightarrow{\hspace{10pt}\tau\hspace{10pt}}\hspace{10pt}}

\newcommand{\petitetautransition}{\petiteatransition{\tau}}
\newcommand{\dtransition}
{\hspace{10pt}\underrightarrow{\hspace{10pt}\delta\hspace{10pt}}\hspace{10pt}}

\newcommand{\petitedtransition}{\petiteatransition{\delta}}
\newcommand{\ptransition}[1]{\hspace{10pt}\longrightarrow_{#1}\hspace{10pt}}
\newcommand{\ptransitionse}[1]{\longrightarrow_{#1}}
\newcommand{\divtransition}{\hspace{10pt}\dashrightarrow\hspace{10pt}}
\newcommand{\divtransitionse}{\dashrightarrow}
\newcommand{\ket}[1]{|#1\rangle}
\newcommand{\bra}[1]{\langle#1|}
\newcommand{\contexte}[4]{<#1, #2 = #3, #4 >}
\newcommand{\scontexte}[4]{/<#1, #2 = #3, #4 >}
\newcommand{\spcontexte}[1]{/ #1}
\newcommand{\contexteqcq}{<s, q = \rho, f >}
\newcommand{\scontexteqcq}{/<s, q = \rho, f >}
\newcommand{\symbcontexteprob}{\boxplus}
\newcommand{\bigsymbcontexteprob}{\mathop\boxplus}
\newcommand{\contexteprob}[5]{\bigsymbcontexteprob_{#1}\!\!<#2, #3 = #4, #5 >}
\newcommand{\contexteprobqcq}{\bigsymbcontexteprob_{p_i}\!\!<s_i, q_i = \rho_i, f_i>}

\newcommand{\spcontexteprob}[2]{/\bigsymbcontexteprob_{#1}#2}
\newcommand{\spcontexteprobbin}[3]{/ #2\symbcontexteprob_{#1}#3}
\newcommand{\contextestable}{\downarrow}
\newcommand{\trace}[1]{Tr(#1)}
\newcommand{\traceout}[3]{Tr_{#1/#2}(#3)}
\newcommand{\varpile}[1]{\mbox{Var$(#1)$}} 
\newcommand{\pileconcat}{|} 
\newcommand{\pileajout}{.}
\newcommand{\pileajoutp}[2]{#2.#1}
\newcommand{\rmpile}[2]{#1\backslash#2}
\newcommand{\dom}[1]{\mbox{dom($#1$)}}
\newcommand{\surcharge}[2]{#1 \cup #2}
\newcommand{\domsoust}[2]{#1\backslash #2}
\newcommand{\codomsoustq}[2]{#1 \backslash #2}
\newcommand{\tailleseq}[1]{\mbox{size}(#1)}
\newcommand{\transfu}{\mathcal U}
\newcommand{\obs}{\mathcal O}
\newcommand{\mstdp}[2]{M_{std, #1}[#2]}
\newcommand{\mstd}[1]{M_{std, #1}}
\newcommand{\mplusmoins}{M_{+-}}

\newcommand{\adjoint}[1]{#1^\dagger}
\newcommand{\appli}[1]{\mathcal{T}_{#1}}
\newcommand{\applip}[2]{\appli{#1}(#2)}
\newcommand{\gate}[1]{\mbox{\it #1}}
\newcommand{\proc}[1]{\mbox{\bf #1}}
\newcommand{\sautdeligne}{\vspace{10pt}}
\newcommand{\tab}{\hspace{20pt}}
\newcommand{\n}{I\!\! N}
\newcommand{\ensemble}[1]{\{#1\}}

\title{ A Process Algebraic Approach to Concurrent and Distributed Quantum Computation:
Operational Semantics }
\author{Marie {\sc Lalire}\footnote{Marie.Lalire@imag.fr} ,
Philippe {\sc Jorrand}\footnote{Philippe.Jorrand@imag.fr}\\
Leibniz Laboratory\\
46, avenue F\'elix Viallet
38000 Grenoble, France}

\begin{document}
\maketitle

\begin{abstract}
Full formal descriptions of algorithms making use of quan\-tum principles must take into account both quan\-tum and classical computing components and assemble them so that they communicate and cooperate.
Moreover, to model concurrent and distributed quan\-tum computations, as well as quan\-tum communication protocols, quan\-tum to quan\-tum communications which move qubits physically from one place to another must also be taken into account.

Inspired by classical process algebras, which provide a framework for modeling cooperating
computations, a process algebraic notation is defined, named QPAlg for Quan\-tum Process Algebra, which provides a homogeneous style to formal descriptions of concurrent and distributed computations comprising both quan\-tum and classical parts.
On the quantum side, QPAlg provides quantum variables, operations on quantum variables (unitary operators and measurement observables), as well as new forms of communications involving the quantum world. The operational semantics makes sure that these quantum objects, operations and communications operate according to the postulates of quantum mechanics.
\end{abstract}

\section{ Introduction }
\label{sectionIntroduction}

Quantum algorithms are frequently described by means of quantum gate networks.
This has several drawbacks, for instance, gate networks do not allow descriptions of loops nor conditional execution of parts of networks. 
So as to overcome these difficulties, a few quantum programming languages have been developed, such as:
QCL \cite{Omer00QCL}, an imperative language designed by Bernhard \"Omer which aims at simulating quantum programs,
qGCL \cite{Zuliani01These} by Paolo Zuliani which allows the construction of proved correct quantum programs through a refinement method, and QPL \cite{Selinger03QPL}, a functional language designed by Peter Selinger with a denotational semantics.
A quantum lamb\-da calculus \cite{Tonder03LambdaCalculus}, based on a simplified linear
lambda calculus, as also been developed by Andr\'e van Tonder.
  
Cooperation between quantum and classical computations is inherent in quan\-tum algorithmics.
For example, the quantum computation part is in general probabilistic:
it produces a result which is checked by a classical part and, if this result is not correct, the quantum computation has to be repeated.
Teleportation of a qubit state from Alice to Bob \cite{BennettBrassard93Teleportation} is another good example of this cooperation. Indeed, Alice carries out a measurement, the result of which (two bits) is sent to Bob, and Bob uses this classical result to determine which quantum transformation he must apply.
Moreover, initial preparation of quantum states and measurement of quantum results are two essential forms of interactions between the classical and quantum kinds of computations which the language must be able to express.
Process algebras are a good candidate for such a language since they provide a framework for modeling cooperating computations.
In addition, they have well defined semantics and permit the transformation of programs as well as the formal study and analysis of their properties.

Process algebras have already been used in the context of quantum programming
in \cite{NagGay02Verif}, where Simon Gay and Rajagopal Nagarajan have modeled a quantum cryptographic protocol and verified its correctness with a classical process algebra. 
Starting with a classical process algebra described in appendix \ref{annexCPA}, this paper explains how to "quantumize" it in section \ref{sectionQuantProcAlg}. Examples of short quantum programs are given in section \ref{sectionExamples}.

\section{ "Quantumized" Processes }
\label{sectionQuantProcAlg}

\subsection{ Quantum Variables }
 \label{subsecQuantVar}

For the purpose of this paper, we consider that there are two types of variables in the
"quan\-tumized" process algebra, one classical: {\it $\nattype$}, for variables taking integer values, and one quan\-tum: {\it $\qubittype$} for variables standing for qubits. An extended version of the process algebra would of course also include quantum registers and other types of variables.

In classical process algebras, variables are instantiated when communications between processes  occur and cannot be modified after their instantiation. As a consequence, it is not necessary to store their values. In fact, when a variable is instantiated, all its occurrences are replaced by the value received (see the semantics of communication in parallel composition, as given in appendix \ref{annexCPA}).

Here, quan\-tum variables stand for physical qubits. Applying a unitary transformation to a variable which represents a qubit modifies the state of that qubit.
This means that values of variables are modified. For that reason, it is necessary to keep track of both variable names and variable states.

Since variables are no longer just names standing for communicated values, they have to be declared. The syntax of declarations is:
$\debutbloc \declvar{x_1}{t_1},\ldots,\declvar{x_n}{t_n} \findeclvar P \finbloc$
 where  $x_1,\ldots,x_n$ is a list of variables, $t_1,\ldots,t_n$ are their types, and $P$ is a process which can make use of these classical and quan\-tum variables.
To simplify the rest of this paper, the names of variables will always be considered distinct. 

In the inference rules which describe the semantics of processes, the states of processes can no longer be process terms only, as was the case for the classical process algebra,
they have to be process terms $P$ together with contexts $C$, of the form $P\spcontexte C$.
The main purpose of a context is to maintain the quan\-tum state,
stored as $q = \rho$ where $q$ is a sequence of quan\-tum variable names
and $\rho$ a density matrix representing their quan\-tum state. 
Moreover, in order to treat classical variables in a similar way, 
modifications of classical variables are also allowed. So, for the same reason as in the case of quan\-tum variables, classical values are stored in the context.
Storing and retrieving classical values is represented by functions
$f: \mbox{\it names} \rightarrow\mbox{\it values}$.
The context must also keep track of the embedding of variable scopes.
To keep track of parallel composition,
this is done via a "cactus stack" structure of sets of variables, called the environment stack ($s$), which stores variable scopes and types. The set of all the variables in $s$ is denoted $\varpile s$,
"$\pileajout$" adds an element on top of a stack, and "$\pileconcat$" concatenates two stacks.

In summary, the context has three components $\contexteqcq$, where:
\begin{itemize}
\item $s$ is the environment stack;
\item $q$ is a sequence of quan\-tum variable names;
\item $\rho$ is a density matrix representing the quan\-tum state of the variables in $q$;
\item $f$ is the function which associates values to classical variables.
\end{itemize}

\sautdeligne
The rules for declaration and liberation of variables are the following:

\sautdeligne
\noindent
{\bf Declaration:}
$$
\frac{}{\debutbloc \declvar{x_1}{t_1},\ldots,\declvar{x_n}{t_n}\findeclvar
P \finbloc \spcontexte{C} \transition \debutbloc P \finbloc \spcontexte{C'}}
$$
with $C = \contexteqcq$, $C' = \contexte{s'}{q}{\rho}{f}$\\
and $s' = \pileajoutp{s} {\ensemble{(x_1,t_1),\ldots,(x_n,t_n)}}$

\sautdeligne
\noindent
This rule adds the new variable names and types on top of the stack $s$. Because the variables do not have values yet, the quan\-tum state and the classical function do not have to be modified at this point.

\sautdeligne
\noindent
{\bf Evolution of a process within the scope of declared variables:}
$$
\frac{P \spcontexte{C} \divtransition P' \spcontexte{C'}}
{\debutbloc P \finbloc \spcontexte{C} \divtransition \debutbloc P'\finbloc \spcontexte{C'}}
$$
where $\divtransitionse$ stands for any of the transitions:
$\petiteatransition\alpha$ with $\alpha$ an action,
$\petitetautransition$ with $\tau$ the "silent" action,  and the declaration transition $\transitionse$.

In short:
if the process $P$ can perform a transition, then the process $\debutbloc P \finbloc$
can perform the same transition, provided that the action of the transition is not $\delta$. 

\sautdeligne
\noindent
{\bf Termination of a process with exit from a scope and liberation of the variables:}
$$
\frac{P \spcontexte{C} \dtransition P' \scontexte{\pileajoutp s e}{q}{\rho}{f}}
{\debutbloc P \finbloc \spcontexte{C} \dtransition \stopproc\ 
\scontexte{s}{\codomsoustq q e}{\traceout{e}{q}{\rho}}{\domsoust{f}{e}}}
$$

If the action is $\delta$, this means that $P$ has successfully terminated, so the context must be cleaned up by eliminating the variables having their scope limited to that process. 
These variables have their names listed in the head $e$ of the stack.
So, cleaning up the context means eliminating the head of the stack, removing the variables in $e$ from the sequence $q$ and from the domain of the function $f$.
The quantum state is obtained by performing a pratial trace on $\rho$ over the qubits in $e$, which is denoted $\traceout{e}{q}{\rho}$.

\subsection{ Basic Actions }

The classical  basic actions are classical to classical communications. Classical to quan\-tum communications and quan\-tum to quan\-tum communications are introduced for respectively initializing qubits and allowing the description of communication protocols. Quan\-tum to classical communications are part of measurement and are dealt with in the next paragraph.

The semantics of communications is based upon the following rules:
$$
\frac{}{g\envoi{v} \prefix P \spcontexte{C} \atransition{g\envoi{v}} P \spcontexte{C}} 
\hspace{10pt} v \in \n 
$$
$$
\frac{}{g\envoi x\prefix P \spcontexte C \atransition{g\envoi x} P \spcontexte{C'}}
$$
where
\begin{itemize}
\item $C = \contexteqcq$ and $C' = \contexte{\rmpile s {\ensemble x}}
{\codomsoustq{q}{\ensemble x}}{\traceout{\ensemble x}{q}{\rho}}{f}$
\item $x\in\varpile s$ and $x\in q$
\end{itemize}
$$
\frac{}{g\recep x \prefix P \spcontexte C \atransition{g\recep x} P \spcontexte C}
$$
with $C=\contexteqcq$, $x\in \varpile s$, and $x\not\in q$. 

\sautdeligne
The first rule deals with classical value sending, the second one, with qubit sending, and the last one, with value reception.
For qubit sending (second rule), because of the no-cloning theorem, the sent qubit must be removed from the context.
It should be noted that in the third rule, the variable $x$ can be classical or quan\-tum
but, if it is quan\-tum, it must not have already been initialized.

In the operational semantics of parallel composition, the combination of these rules defines communication.
In a classical to quan\-tum communication, the qubit is initialized in the basis state $\ket v \bra v$, where $v$ is the classical value sent (in this case, $v$ must be $0$ or $1$). In a quantum to quantum communication, the name of the sent qubit is replaced in $q$ by the name of the receiving qubit.

\sautdeligne
The second kind of basic actions is unitary transformations which perform the unitary evolution of qubit states. Given a set $\transfu$ of predefined unitary transformations, the action corresponding to the application of $U\in\transfu$ to a list of quan\-tum variables is denoted by $U[x_1,\ldots,x_n]$. 

The inference rule for unitary transformations is:
$$
\frac{}{U[x_1,\ldots,x_n]\prefix P \scontexteqcq \tautransition
P \scontexte{s}{q}{\rho'}{f}}
$$
where
\begin{itemize}
\item  $U\in \transfu$, $x_1,\ldots,x_n \in \varpile s$, and $x_1,\ldots,x_n \in q$ 
\item $\forall\ i,j \in \ensemble{0,\ldots,n}$ such that $i \neq j$ : $\ x_i \neq x_j$
\item $\rho' = \applip{U}{\rho}$
\end{itemize}

The condition $x_1,\ldots,x_n \in q$ prevents from applying a unitary transformation to qubits which have not been initialized.

In the third point, $\appli{U}$ is the super-operator which must be applied to $\rho$, to describe the evolution of the quantum state due to the application of the unitary transformation $U$ to the qubits $x_1,\ldots,x_n$. In general, with $A$ a matrix:

$$\appli{A}: \rho \mapsto\adjoint{\Pi}.(A\otimes I^{\otimes k}).\Pi.\rho.\adjoint{\Pi}.
(\adjoint{A}\otimes I^{\otimes k}).\Pi$$
where
\begin{itemize}
\item $\Pi$ is the permutation matrix which places the $x_i$'s at the head of $q$
\item $k = \tailleseq q - n\ $
\item $I^{\otimes k}\!= \underbrace{I \otimes \cdots \otimes I}_k$, where $I$ is the identity matrix on $\mathbb C^2$
\end{itemize}

Since the unitary transformation $U$ may be applied to qubits which are anywhere within the list $q$, a permutation $\Pi$ must be applied first. This permutation moves the $x_i$'s so that they are placed at the head of $q$ in the order specified by $[x_1,\ldots,x_n]$.
Then $U$ can be applied to the first $n$ elements and $I$ to the remainder. Finally, the last operation is the inverse of the permutation $\Pi$
so that at the end, the elements in $q$ and $\rho$ are placed back in the same order.

\subsection{ Measurement and Probabilistic Processes }

A last but essential basic action has to be introduced into the process algebra: quan\-tum measurement.
Let $M\in\obs$ be an observable, $x_1,\ldots,x_n$ a list of distinct quan\-tum variables and $g$ a gate. Then, the syntax for measurement is the following:
\begin{itemize}
\item $M[x_1,\ldots,x_n]$ is a measurement of the $n$ qubits of the list with respect to observable $M$, but the classical result is neither stored nor transmitted.
\item $g \envoi{M[x_1,\ldots,x_n]}$ is a measurement of the $n$ qubits of the list with respect to observable $M$, followed by sending the classical result through gate $g$.
\end{itemize}

Measurement is probabilistic: more precisely, the classical result and the quan\-tum state after measurement are probabilistic.
In the case of measurement without communication of the classical result, only the quantum state is probabilistic after measurement, so the probabilities can be included in the density matrix.
In the other case, the classical result is probabilistic, this requires the introduction of a probabilistic composition operator for contexts.
This operator is denoted $\symbcontexteprob_p$:
the state $P\spcontexteprobbin{p}{C_1}{C_2}$ is $P\spcontexte{C_1}$ with probability $p$ and $P\spcontexte{C_2}$ with probability $1-p$.

This implies that, in general, the context is either of the form $\contexteqcq$, or of the form
$\contexteprobqcq$ where the $p_i$'s are probabilities adding to $1$.

As explained in \cite{Cazorla01Art} and \cite{CazorlaMis}, if a process contains both a probabilistic and a nondeterministic choice, the probabilistic choice must always be solved first.
In the process algebra presented here, nondeterminism appears with parallel composition and conditional choice. So as to guarantee that probabilistic choice is always solved first, the notion of probabilistic stability for contexts is introduced: a context $C$ is probabilistically stable, which is denoted $C\contextestable$, if it is of the form $\contexteqcq$.
If the context of a process state is not stable, a probabilistic transition must be performed first.

The semantic rule for measurement without communication is:
$$
\frac{}{M[x_1,\ldots,x_n]\prefix P \scontexteqcq \tautransition
P\scontexte{s}{q}{\rho'}{f}}
$$
with
\begin{itemize}
\item $x_1,\ldots,x_n \in \varpile s$ and $x_1,\ldots,x_n \in q$ 
\item $\forall\ i,j \in \ensemble{0,\ldots,n}$ such that $i \neq j$ : $\ x_i \neq x_j$
\item $M\in\obs$  with $\sum_{i} \lambda_i P_i$ as spectral decomposition
\item $\rho' =  \sum_{i}\applip{P_i}{\rho}$
\end{itemize}

As in the case of unitary transformations, $\appli{P_i}$ is the super-operator corresponding to the application of the projector $P_i$ to measured qubits.
The computation of $\rho'$ stems from the projective measurement postulate of quan\-tum mechanics.

When the value coming out of the measurement is sent out, the rule is:
$$
\frac{}{g\envoi{M[x_1,\ldots,x_n]} \prefix P \spcontexte C \tautransition
\debutbloc g\envoi y\prefix \term \finbloc \seq P\spcontexteprob{p_i}{C_i}}
$$
where
\begin{itemize}
\item $C = \contexteqcq$ (which implies $C\contextestable$)
\item $C_i = \contexte{\pileajoutp s {\ensemble{(y,\nattype)}}}{q}{\rho_i}
{\surcharge f \ensemble{y\mapsto\lambda_i}}$
\item $x_1,\ldots,x_n \in \varpile s$ and $x_1,\ldots,x_n \in q$ 
\item $\forall\ i,j \in \ensemble{0,\ldots,n}$ such that $i \neq j$ : $\ x_i \neq x_j$
\item $y$ is a new variable (implicitly declared as $\cvar y$, see below)
\item $M\in\obs$  with $\sum_{i} \lambda_i P_i$ as spectral decomposition
\item $p_i = \trace{\applip{P_i}{\rho}}$
\item $\rho_i = \frac{1}{p_i}\applip{P_i}{\rho}$
\end{itemize}

\sautdeligne
The only remaining point is the evolution of processes within probabilistic contexts. It is necessary to introduce probabilistic transitions for describing this evolution:
$$S_1 \ptransition p S_2$$
means that state $S_1$ becomes $S_2$ with probability $p$.
This is used in the following rule:
$$
\frac{}{P\spcontexteprob{p_i}{C_i} \ptransition{p_i} P\spcontexte{C_i}}
\mbox{ where } \sum_j p_j = 1
$$

The syntax and the main inference rules of this quan\-tum process algebra are presented in appendix \ref{annexQPA}.


\section{ Examples }
\label{sectionExamples}

\newcommand{\eprinit}{\proc{BuildEPR}}
\newcommand{\verifepr}{\proc{CheckEPR}}
\newcommand{\alice}{\proc{Alice}}
\newcommand{\bob}{\proc{Bob}}
\newcommand{\teleport}{\proc{Teleport}}
\newcommand{\meas}{\gate{meas}}
\newcommand{\eve}{\proc{Eve}}
\newcommand{\canal}{\proc{Channel}}
\newcommand{\faille}{\proc{Flaw}}
\newcommand{\prot}{\proc{Protocol}}
\newcommand{\aaa}{\proc{A}}
\newcommand{\bbb}{\proc{B}}
\newcommand{\eee}{\proc{E}}
\newcommand{\fillcanal}{\gate{fill}}
\newcommand{\emptycanal}{\gate{empty}}
\newcommand{\fillfaille}{\gate{fillFlaw}}
\newcommand{\emptyfaille}{\gate{emptyFlaw}}
\newcommand{\random}{\proc{Random}}
\newcommand{\aaaun}{\proc{A$_1$}}
\newcommand{\aaadeux}{\proc{A$_2$}}
\newcommand{\dataa}{\gate{dataA}}
\newcommand{\datab}{\gate{dataB}}
\newcommand{\basea}{\gate{baseA}}
\newcommand{\baseb}{\gate{baseB}}
\newcommand{\bool}{\gate{bool}}
\newcommand{\ok}{\gate{ok}}
\newcommand{\keepdataa}{\gate{keepDataA}}
\newcommand{\keepdatab}{\gate{keepDataB}}
\newcommand{\base}{\gate{base}}
\newcommand{\keep}{\gate{keep}}
\newcommand{\received}{\gate{received}}

In the following examples, the set $\transfu$ of unitary transformations is:
$$\transfu = \{H, CNot, I, X, Y, Z\}$$
where $H$ is Hadamard transformation, $CNot$ is the "controlled not" operation, $I$ is the identity, and $X, Y, Z$ are Pauli matrices.
The set $\obs$ of observables contains the observables corresponding to measurement of one and two qubits in the standard basis, denoted respectively $\mstd{1}$ and $\mstd{2}$, and the observable corresponding to measurement of a qubit in the basis $\ensemble{\ket{+},\ket{-}}$, denoted $\mplusmoins$.

\subsection{Construction of an EPR pair}

$$
\begin{array}{rcl}
\eprinit & \procdef & \debutbloc \qvar x, \qvar y \findeclvar \\
&& \tab (( g_1\recep{x} \prefix g_2 \recep y \prefix H[x] \prefix CNot[x,y] \prefix\term)\\
&& \tab \para (g_1\envoi 0\prefix g_2\envoi 0\prefix\term)) \restrictg{g_1,g_2}\\
&& \finblocse \\
\end{array}
$$

This process puts the pair of qubits $x, y$ in the state
$\ket{EPR}= \frac{1}{\sqrt 2}(\ket{00}+\ket{11})$.
To check that the order of measurement of the two qubits does not matter, it is possible, using the inference rules, to analyze the behavior of the following two processes: in both of them, the first measurement produces $0$ ($1$) with probability $0.5$ and the second measurement produces
$0$ ($1$) with probability 1.

$$
\begin{array}{rcl}
\verifepr_{\bf 1} &\procdef& \debutbloc \qvar a, \qvar b \findeclvar\\
&& \tab \eprinit[a,b]\seq \mstdp{1}{a} \prefix \mstdp{1}{b} \prefix \term\\
&& \finblocse\\
&&\\
\verifepr_{\bf 2} &\procdef& \debutbloc \qvar a, \qvar b \findeclvar\\
&& \tab \eprinit[a,b] \seq \mstdp{1}{b} \prefix \mstdp{1}{a} \prefix \term \\
&& \finblocse\\
\end{array}
$$

\subsection{Teleportation}

Once upon a time, there were two friends, Alice and Bob who had to separate and live away from each other.
Before leaving, each one took a qubit of the same EPR pair.
Then Bob went very far away, to a place that Alice did not know.
Later on, someone gave Alice a mysterious qubit in a state $\ket\psi = \alpha\ket 0+\beta\ket 1$,
with a mission to forward this state to Bob.
Alice could neither meet Bob and give him the qubit, nor clone it and broadcast copies everywhere, nor obtain information about $\alpha$ and $\beta$. Nevertheless, Alice succeeded thanks to the EPR pair and the teleportation protocol \cite{BennettBrassard93Teleportation}:

\sautdeligne
$\begin{array}{lcl}
\alice&\procdef& \debutbloc \qvar x, \qvar y \findeclvar\\
&& \tab CNot[x,y] \prefix H[x] \prefix \meas \envoi{\mstdp{2}{x,y}} \prefix \term\\
&& \finblocse \\
\end{array}$

\sautdeligne
$\begin{array}{lcl}
\bob&\procdef&\debutbloc \qvar z \findeclvar\\
&& \tab \debutbloc \cvar k \findeclvar\\
&& \tab\tab \meas\recep k\prefix\\
&& \tab\tab \debutbloc \cond{k=0}{I[z]\prefix\term}, \\
&& \tab\tab \cond{k=1}{X[z]\prefix\term}, \\
&& \tab\tab \cond{k=2}{Z[z]\prefix\term}, \\
&& \tab\tab \cond{k=3}{Y[z]\prefix\term} \finbloc \\
&&\tab \finblocse\\
&& \finblocse \\
\end{array}$

\sautdeligne
$\begin{array}{lcl}
\teleport&\procdef& \debutbloc \qvar\psi \findeclvar\\
&& \tab \debutbloc \qvar a, \qvar b \findeclvar\\
&& \tab\tab \eprinit[a,b]\seq\\
&& \tab\tab (\alice[\psi,a]\para\bob[b])\restrictg\meas\\
&& \tab \finblocse\\
&& \finblocse\\
\end{array}$

\sautdeligne
The inference rules can be used to show that this protocol results in Bob's $z$ qubit having the state initially possessed by the $x$ qubit of Alice, with only two classical bits sent from Alice to Bob.

\subsection{Communication protocols}

Alice sends qubits to Bob through a non secure channel and Eve eavesdrops this channel to get information on the qubits sent by Alice.
In the following example $\aaa$, $\bbb$, and $\eee$ are processes modeling whatever Alice, Bob, and Eve may respectively apply to their qubits. The actions of these processes, which are not made explicit here, will be specified in the next example of the BB84 protocol.

The communication protocols which are described here could be used to mo\-del cryptographic protocols so as to check if they are secure.

\sautdeligne
\subsubsection*{Eve intercepts all qubits}

Eve intercepts qubits because of a flaw in the channel that Alice and Bob use to communicate.

\sautdeligne
$\begin{array}{lcl}
\alice & \procdef & \debutbloc \qvar a \findeclvar
\aaa[a]\seq \fillcanal \envoi a \prefix\term \finbloc \seq \alice\\
\end{array}$

\sautdeligne
$\begin{array}{lcl}
\bob &\procdef& \debutbloc \qvar b \findeclvar
\emptycanal \recep b \prefix \bbb[b] \finbloc \seq \bob\\
\end{array}$

\sautdeligne
$\begin{array}{lcl}
\eve &\procdef& \debutbloc \qvar e, \qvar f \findeclvar\\
&& \tab \emptyfaille \recep e \prefix \eee[e,f] \seq \fillfaille \envoi f \prefix \term\\
&& \finblocse \seq \eve\\
\end{array}$

\sautdeligne
$\begin{array}{lcl}
\faille &\procdef& \debutbloc \qvar u, \qvar v\findeclvar
\emptyfaille\envoi u \prefix \fillfaille \recep v \prefix\term \finbloc\\
\end{array}$

\sautdeligne
$\begin{array}{lcl}
\canal &\procdef& \debutbloc \qvar x, \qvar y\findeclvar
\fillcanal \recep x \prefix\faille[x,y] \seq \emptycanal\envoi y \prefix\term \finbloc \seq\\
&&\canal\\
\end{array}$

\sautdeligne
$\begin{array}{lcl}
\prot &\procdef& (\alice\para\bob\para\eve\para\canal)\\
&& \tab\restrict \{\fillcanal,\emptycanal,\fillfaille,\emptyfaille\}
\end{array}$

\sautdeligne
\subsubsection*{Eve intercepts some of the qubits}

This part requires that a nondeterministic process composition $P\nondet Q$ be introduced in the process algebra. This can be done, provided that probabilistic choices are always solved first (this operator is not presented in the operational semantics in appendix \ref{annexQPA}).

To model the fact that Eve does not succeed in intercepting all qubits, the flaw in the channel is made nondeterministic:
$$
\begin{array}{lcl}
\canal &\procdef& \debutbloc \qvar x \findeclvar\\
&& \tab \fillcanal \recep x \prefix\\
&& \tab (\\
&& \tab\tab \debutbloc \qvar y \findeclvar \faille[x,y]\seq\emptycanal\envoi y\prefix\term \finbloc
\nondet\\
&& \tab\tab (\emptycanal\envoi x \prefix\term)\\
&& \tab )\\
&& \finblocse \seq \canal
\end{array}
$$


\subsection{The BB84 Protocol}

The BB84 protocol \cite{BennettBrassard84BB84} is a protocol of quan\-tum key distribution: Alice and Bob must agree on a private key, i.e. a list of bits that should remain secret. To communicate, they can send qubits through a non secure channel. In fact, the processes $\aaa$ and $\bbb$ left unspecified in the previous paragraph can be used to model this protocol. 
The process $\alice$ is redefined and the process $\bbb$ used by $\bob$ is made explicit.
In addition, another process is defined: the process $\random$ which initializes a bit randomly at $0$ or $1$.
The gates $\keepdataa$ and $\keepdatab$ are used by Alice and Bob respectively to send the bits that they want to keep.
In this example, we take the liberty of using identical names for variables having distinct scopes.

\sautdeligne
$\begin{array}{lcl}
\alice & \procdef & \debutbloc \qvar{a}, \cvar{\dataa}, \cvar{\basea} \findeclvar\\
&&\tab \aaaun[a,\dataa,\basea] \seq \fillcanal \envoi a \prefix  \aaadeux[\dataa,\basea]\\
&&\finblocse \seq \alice
\end{array}$

\sautdeligne
$\begin{array}{lcl}
\random &\procdef& \debutbloc \cvar r \findeclvar\\
&& \tab \debutbloc \qvar x \findeclvar\\
&& \tab\tab (g\envoi 0\prefix\term \para g\recep x\prefix\term)\restrictg{g} \seq\\
&& \tab\tab H[x]\prefix\\
&& \tab\tab (h \envoi{\mstdp 1 x}\prefix\term \para h\recep r\prefix\term)\restrictg{h}\\
&& \tab \finblocse \\
&& \finblocse \\
\end{array}$

\sautdeligne
$\begin{array}{lcl}
\aaaun & \procdef & \debutbloc \qvar a, \cvar \dataa, \cvar \basea \findeclvar\\
&& \tab \random[\dataa][a]\seq\\
&& \tab \random[\basea]\seq\\
&& \tab \debutbloc \cond{\basea=1}{H[a]\prefix\term} \finbloc \\
&&  \finblocse \\
\end{array}$

\sautdeligne
$\begin{array}{lcl}
\aaadeux & \procdef & \debutbloc \cvar\dataa,\cvar\basea \findeclvar\\
&& \tab \debutbloc \cvar {\bool}, \cvar\ok \findeclvar \\
&& \tab\tab \received \recep \ok \prefix\\
&& \tab\tab \base \envoi \basea \prefix\\
&& \tab\tab \keep \recep \bool \prefix\\
&& \tab\tab \debutbloc \cond{\bool=1}{\keepdataa \envoi \dataa \prefix\term} \finbloc \\
&& \tab \finblocse \\
&& \finblocse \\
\end{array}$

\sautdeligne
$\begin{array}{lcl}
\bbb &\procdef& \debutbloc \qvar b \findeclvar\\
&& \tab \debutbloc \cvar\basea, \cvar\baseb, \cvar\datab \findeclvar\\
&& \tab\tab \random[\baseb]\seq\\
&& \tab\tab (\\
&& \tab\tab\tab\debutbloc \cond{\baseb=0}{g\envoi{\mstdp 1 b}\prefix\term},\\
&& \tab\tab\tab\ \ \cond{\baseb=1}{g\envoi{M_{+-}[b]}\prefix\term} \finbloc \\
&& \tab\tab\tab \para g \recep \datab \prefix \term\\
&& \tab\tab )\restrictg g\seq\\

&& \tab\tab \received \envoi 1 \prefix\\
&& \tab\tab \base \recep \basea \prefix\\
&& \tab\tab \debutbloc
\cond{\basea=\baseb}{\keep \envoi 1\prefix\keepdatab \envoi{\datab}\prefix \term},\\
&& \tab\tab \ \ \cond{\basea\neq\baseb}{\keep \envoi 0 \prefix \term} \finbloc\\
&& \tab \finblocse\\
&& \finblocse\\
\end{array}$

\section{ Conclusion }

This paper has presented a process algebra for quantum programming.  One of its advantages is that it can describe classical and quantum programming, and their cooperation. Without this cooperation, the implementation of the above protocols is not possible. Another feature of this language is that measurement and initialization of quantum registers appear through communications between quantum and classical parts of the language, which happens to be a faithful model of physical reality.

Moreover, a thorough semantics has been defined, thus allowing the study and analysis of programs. One peculiarity of  this  semantics is the introduction of probabilistic processes, due to quantum measurement. Probabilistic processes perform probabilistic transitions. As a consequence, the execution tree obtained  from a process presents action branches and probabilistic branches.

Several extensions are possible. As already mentioned, a nondeterministic process composition operator can be introduced. A probabilistic composition of processes could be added. This would allow, for example, the description of communication protocols in which Eve intercepts qubits with a given probability.
Another direction of study concerns the definition of an equivalence among processes, which is necessary for obtaining a more abstract semantics.




\appendix

\section{A Classical Process Algebra}
\label{annexCPA}

The classical process algebra chosen here is quite similar to CCS \cite{Milner89CommConc} and Lotos \cite{Bolognesi87IsoLotos}.
In this process algebra, communication among processes is the only basic action.
There is a distinction between value emission denoted
 $g \envoi v$, where $g$ is a communication  gate and $v$ a value, and value reception denoted $g \recep x$, where $g$ is a gate and $x$ a variable which receives the value.
To create a process from basic actions, the prefix operator "$\prefix$" is used: if $\alpha$ is an action and $P$, a process, $\alpha\prefix P$ is a new process which performs $\alpha$ first, then behaves as $P$.

There are two predefined processes. The first one is $\stopproc$, the process that cannot perform any transition, and the other one is $\term$, which performs a "$\delta$-transition" for signaling successful termination, and becomes $\stopproc$ ("$\delta$-transitions" are necessary in the semantics of sequential composition of processes).

The operators of the process algebra are: sequential composition ($P\seq Q$), parallel composition ($P\para Q$), conditional choice
($\debutbloc \cond{c_1}{P_1},\ldots,\cond{c_n}{P_n}\finbloc$)
and restriction ($P\restrict L$). As for sequential composition, process $Q$ is executed if process $P$ terminates successfully, that is to say if $P$ performs a $\delta$-transition.
The process $\debutbloc \cond{c_1}{P_1},\ldots,\cond{c_n}{P_n}\finbloc$, where $c_i$ is a condition and $P_i$ a process, evolves as a process chosen nondeterministically among the processes $P_j$ such that $c_j$ is true.
Restriction is useful for disallowing the use of some gates (the gates listed in $L$),
thus forcing internal communication within process $P$. 
Communication can occur between two parallel processes whenever a value emission in one of them and a value reception in the other one use the same gate name. For instance, a communication can occur on gate $g$ in the process $g\envoi v\prefix P\para g\recep x \prefix Q$. After the communication has occurred, this process  becomes $P\para Q[x\leftarrow v]$ where $Q[x\leftarrow v]$ is $Q$ where all occurrences of $x$ have been replaced by $v$.

\subsection{Syntax of Process Terms}

\begin{supertabular}{lcl}
\it process & $\syntaxdef$ & \it $\pmb\stopproc$\\
                  &$|$& \it $\pmb\term$ \\
                  &$|$& \it communication $\pmb\prefix$ process\\
                  &$|$& \it  process $\pmb\seq$ process\\
                  &$|$& \it  process $\pmb\para$ process\\
                  &$|$& \it  process $\pmb \backslash \pmb\{$ gate\_list $\pmb\}$ \\
                  &$|$& \it  $\pmb\debutbloc$ cond\_list $\pmb\finbloc$\\
                  &$|$& \it process\_name \\
&&\\
\it communication & $\syntaxdef$ & \it gate {\bf !} exp $|$ gate {\bf ?} variable  \\
&&\\
\it cond & $\syntaxdef$ & \it bexp $\pmb\rightarrow$ process  \\
&&\\
\it proc\_def & $\syntaxdef$ & \it process\_name $\displaystyle\pmb{\procdef}$ process\\
\end{supertabular}

\subsection{Semantics}
\label{annexSemCPA}

The semantics is specified with inference rules which give the evolution of the states of processes. In the classical process algebra considered here, the state of a process is a process term, and
there are three kinds of transitions:
\begin{itemize}
\item action transition: $\petiteatransition\alpha$ where $\alpha$ is $g\envoi v$ or $g\recep x$;
\item silent transition: $\petitetautransition$, for internal transition;
\item delta transition: $\petitedtransition$, for successful termination.
\end{itemize}

In the following, $P, Q, P', Q', P_i$ and $P_i'$ are processes, $\alpha$ and $\alpha_i$ are actions,
$g$ is a communication gate, $v$ is a value, $x$ is a variable, and $c_j$ is a condition.

\sautdeligne
\subsubsection*{Successful termination}
$$
\frac{}{\term \dtransition \stopproc}
$$

\subsubsection*{Action Prefix}
\begin{tabular}{p{0.46\linewidth}p{0.46\linewidth}}
$\displaystyle \frac{}{g \envoi v \prefix P \atransition{g\envoi v} P}\hspace{10pt}v\in\n$
&
$\displaystyle \frac{}{g \recep x \prefix P \atransition{g\recep x} P}$
\end{tabular}

\subsubsection*{Sequential composition}
\begin{tabular}{p{0.46\linewidth}p{0.46\linewidth}}
$\displaystyle \frac{P \atransition\alpha P'}{P\seq Q \atransition\alpha P'\seq Q}
\hspace{10pt} \alpha \neq \delta$
&
$\displaystyle \frac{P \dtransition P'}{P\seq Q \tautransition Q}$
\end{tabular}

\subsubsection*{Parallel composition}
\begin{tabular}{p{0.46\linewidth}p{0.46\linewidth}}
$\displaystyle \frac{P \atransition\alpha P'}{P\para Q \atransition\alpha P'\para Q}
\hspace{10pt} \alpha \neq \delta$
&
$\displaystyle \frac{Q \atransition\alpha Q'}{P\para Q \atransition\alpha P\para Q'}
\hspace{10pt} \alpha \neq \delta$
\end{tabular}
$$
\frac{P\atransition{g\envoi v}P' \hspace{10pt} Q\atransition{g\recep x}Q'}
{P\para Q \tautransition P'\para Q'[x\leftarrow v]}
$$
$$
\frac{P\atransition{g\recep x}P' \hspace{10pt} Q\atransition{g\envoi v}Q'}
{P\para Q \tautransition P'[x\leftarrow v]\para Q'}
$$
$$
\frac{P\dtransition P'\hspace{20pt} Q\dtransition Q'}
{P\para Q \dtransition \stopproc}
$$


\subsubsection*{Conditional choice}
$$
\frac{P_i \atransition{\alpha_i} P_i'}
{\debutbloc \cond{c_1}{P_1},\ldots,\cond{c_n}{P_n} \finbloc \atransition{\alpha_i} P_i' }
\hspace{10 pt} c_i
$$
$$
\frac{}{\debutbloc \cond{c_1}{P_1},\ldots,\cond{c_n}{P_n} \finbloc \dtransition \stopproc }
\hspace{10 pt} \forall i, \neg c_i
$$
\subsubsection*{Restriction}
$$
\frac{P\atransition{\alpha}P'}{P\restrict L \atransition\alpha P'\restrict L}
\hspace{10pt}
\begin{array}{c}
\alpha=\tau \vee \alpha = \delta\\
\vee (\alpha = g[\envoi v \mbox{or}\recep x] \wedge g\not\in L)
\end{array}
$$

\section{The Quantum Process Algebra}
\label{annexQPA}

\subsection{Syntax}

\begin{supertabular}{lcl}
\it process & $\syntaxdef$ & \it $\pmb\stopproc$\\
                  &$|$& \it $\pmb\term$ \\
                  &$|$& \it action $\pmb\prefix$ process\\
                  &$|$& \it  process $\pmb\seq$ process\\
                  &$|$& \it  process $\pmb\para$ process\\
                  &$|$& \it  process $\pmb \backslash \pmb\{$ gate\_list $\pmb\}$ \\
                  &$|$& \it  $\pmb\debutbloc$ cond\_list $\pmb \finbloc$\\
                  &$|$& \it $\pmb \debutbloc$ var\_decl\_list $\pmb \findeclvar$ process $\pmb\finbloc$ \\
                  &$|$& \it process\_name $[\pmb [$ var\_list $\pmb ]]$\\
&&\\
\it action & $\syntaxdef$ & \it communication \\
		&$|$&\it unit\_transf\\
		&$|$&\it measurement \\
&&\\
\it communication & $\syntaxdef$ & \it gate {\bf !} exp\\
                  &$|$&\it gate {\bf !} measurement\\
 		&$|$& \it gate {\bf ?} variable  \\
&&\\
\it unit\_transf         & $\syntaxdef$ & \it unitary\_operator $\pmb [$ var\_list $\pmb ]$ \\
&&\\
\it measurement    & $\syntaxdef$ & \it observable $\pmb [$ var\_list $\pmb ]$\\
&&\\
\it var\_decl & $\syntaxdef$ & \it variable $\pmb :$ var\_type\\
&&\\
\it proc\_def & $\syntaxdef$ & \it process\_name $\displaystyle\pmb{\procdef}$ process\\
\end{supertabular}

\subsection{Main Inference Rules of the Semantics}

With respect to appendix \ref{annexSemCPA}, two new kinds of transitions have been added:
\begin{itemize}
\item declaration transition: $\transitionse$, for variable declaration;
\item probabilistic transition: $\ptransitionse p$, where $p$ is a probability.
\end{itemize}

In the following, $C$, $C'$ or $C_i$ are contexts. 

\subsubsection*{Successful termination}
$$
\frac{}{\term\spcontexte C \dtransition \stopproc\spcontexte C}
\hspace{10pt}C\contextestable
$$
\subsubsection*{Action Prefix}
$$
\frac{}{g \envoi v \prefix P \spcontexte C \atransition{g\envoi v} P \spcontexte C}
\hspace{10pt} v \in \n,\ C\contextestable
$$
$$
\frac{}{g\envoi{x} \prefix P \spcontexte C \atransition{g\envoi{f(x)}} P \spcontexte C} 
$$
where $C=\contexteqcq$, $x \in \varpile s$ and $x \in \dom f$
$$
\frac{}{g\envoi x\prefix P \spcontexte C \atransition{g\envoi x} P \spcontexte{C'}}
$$
where
\begin{itemize}
\item $C = \contexteqcq$ and
$C' = \contexte{\rmpile s {\ensemble x}}{\codomsoustq{q}{\ensemble x}}
{\traceout{q}{\ensemble x}{\rho}}{f}$
\item $x\in\varpile s$ and $x\in q$
\end{itemize}
$$
\frac{}{g \recep x\prefix P \spcontexte C \atransition{g\recep x} P\spcontexte C}
$$
where $C=\contexteqcq$, $x \in \varpile s$

$$
\frac{}{U[x_1,\ldots,x_n]\prefix P \spcontexte C \tautransition
P \spcontexte{C'}}
$$
where
\begin{itemize}
\item $C=\contexteqcq$, $C'=\contexte{s}{q}{\rho'}{f}$
\item  $U\in \transfu$, $x_1,\ldots,x_n \in \varpile s$, and $x_1,\ldots,x_n \in q$ 
\item $\forall\ i,j \in \ensemble{0,\ldots,n}$ such that $i \neq j$ : $\ x_i \neq x_j$
\item $\rho' = \applip{U}{\rho}$
\end{itemize}

\sautdeligne
$\appli{\ }$ is defined in the following way: if $A$ is a matrix, then
$$\appli{A}: \rho \mapsto\adjoint{\Pi}.(A\otimes I^{\otimes k}).\Pi.\rho.\adjoint{\Pi}.
(\adjoint{A}\otimes I^{\otimes k}).\Pi$$
where $\Pi$ is the permutation matrix which places the $x_i$'s at the head of $q$, and
$k = \tailleseq q - n\ $.

$$
\frac{}{M[x_1,\ldots,x_n]\prefix P \scontexteqcq C \tautransition
P\scontexte{s}{q}{\rho}{f}}
$$
with
\begin{itemize}
\item $x_1,\ldots,x_n \in \varpile s$ and $x_1,\ldots,x_n \in q$ 
\item $\forall\ i,j \in \ensemble{0,\ldots,n}$ such that $i \neq j$ : $\ x_i \neq x_j$
\item  $M\in\obs$  with $\sum_{i} \lambda_i P_i$ as spectral decomposition
\item $\rho' = \sum_i \applip{P_i}{\rho}$
\end{itemize}
$$
\frac{}{g\envoi{M[x_1,\ldots,x_n]} \prefix P \spcontexte C \tautransition
\debutbloc g\envoi y\prefix \term \finbloc \seq P\spcontexteprob{p_i}{C_i}}
$$
where
\begin{itemize}
\item $C = \contexteqcq$ (which implies $C\contextestable$)
\item $C_i = \contexte{\pileajoutp s {\ensemble{(y,\nattype)}}}{q}{\rho_i}
{\surcharge f \ensemble{y\mapsto\lambda_i}}$
\item $x_1,\ldots,x_n \in \varpile s$ and $x_1,\ldots,x_n \in q$ 
\item $\forall\ i,j \in \ensemble{0,\ldots,n}$ such that $i \neq j$ : $\ x_i \neq x_j$
\item $y$ is a new variable (implicitly declared as $\cvar y$, see below)
\item $M\in\obs$  with $\sum_{i} \lambda_i P_i$ as spectral decomposition
\item $p_i = \trace{\applip{P_i}{\rho}}$
\item $\rho_i = \frac{1}{p_i}\applip{P_i}{\rho}$
\end{itemize}

\subsubsection*{Probabilistic contexts}
$$
\frac{}{P\spcontexteprob{p_i}{C_i} \ptransition{p_i} P\spcontexte{C_i}}
\mbox{ where } \sum_j p_j = 1
$$

\subsubsection*{Sequential composition}
$$
\frac{P \spcontexte{C} \divtransition P' \spcontexte{C'}}
{P \seq Q \spcontexte{C} \divtransition P'\seq Q \spcontexte{C'}}
$$
where $\divtransitionse$ stands for any of the transitions~: $\petiteatransition\alpha$ with $\alpha$ an action different from $\delta$, $\petitetautransition$, or $\transitionse$.
$$
\frac{P \spcontexte{C} \dtransition P' \spcontexte{C'}}{P \seq Q\spcontexte{C} \tautransition Q \spcontexte{C'}}
$$
\subsubsection*{Parallel composition}

In the rules for parallel composition, $C$, $C_P$ and $C_Q$ are defined as:
\begin{itemize}
\item $C = \contexte{\pileajoutp s {(s_P\para s_Q)}}{q}{\rho}{f}$
\item $C_P = \contexte{s_P\pileconcat s}{q}{\rho}{f}$
\item $C_Q = \contexte{s_Q\pileconcat s}{q}{\rho}{f}$
\end{itemize}

In the definition of $C$, the operator $\para$ permits to build a cactus stack (see paragraph \ref{subsecQuantVar}).
In the cactus stack $\pileajoutp s {(s_P\para s_Q)}$ of the process $P\para Q$, the names in $s$
correspond to variables shared by $P$ and $Q$ whereas the names in $s_P$ (resp. $s_Q$) correspond to variables declared in $P$ (resp. $Q$).
$$
\frac{P \spcontexte{C_P} \divtransition P' \spcontexte{C_P'}}
            {P \para Q \spcontexte{C} \divtransition
              P' \para Q \spcontexte{C'}}
$$
where
\begin{itemize}
\item $\divtransitionse$ stands for one of those transitions~: $\petiteatransition\alpha$ with $\alpha$ an action and $\alpha \neq \delta$, $\petitetautransition$, $\transitionse$

\item If $C_P' =\contexte{s'}{q'}{\rho'}{f'}$ then 
$C' = \contexte{\pileajoutp s {(s_P'\para s_Q)}}{q'}{\rho'}{f'}$
with $s_P'$ such that $s' = s_P'\pileconcat s$ ($P$ can neither add to nor remove variables from $s$)

\item If $C_P' =\contexteprob{p_i}{s_i'}{q_i'}{\rho_i'}{f_i'}$
then $C' = \contexteprob{p_i}{\pileajoutp s {({s_P}_i'\para s_Q)}}{q_i'}{\rho_i'}{f_i'}$
with ${s_P}_i'$ such that $s_i' = {s_P}_i'\pileconcat s$
\end{itemize}

$$
\frac{P \spcontexte{C_P} \atransition{g\envoi v} P' \spcontexte{C_P'}
             \hspace{15pt}
              Q \spcontexte{C_Q} \atransition{g\recep x} Q' \spcontexte{C_Q'}}
            {P \para Q \spcontexte{C} \tautransition P' \para Q' \spcontexte{C'}}
$$
where
\begin{itemize}
\item $x \in \varpile s \cup \varpile{s_Q}$ and $v \in \n$
\item  If $x$ is of type $\nattype$, then:
$C' = \contexte{\pileajoutp s {(s_P\para s_Q)}} {q}{\rho}{\surcharge f \ensemble{x\mapsto v}}$
\item  If $x$ is of type $\qubittype$, then:
$x\not\in q$, $v\in\ensemble{0,1}$\\
and $C' = \contexte{\pileajoutp s {(s_P\para s_Q)}} {x.q}{\ket v\bra v\otimes\rho}{f}$
\end{itemize}

$$
\frac{P \spcontexte{C_P} \atransition{g\envoi x} P' \spcontexte{C_P'}
             \hspace{15pt}
              Q \spcontexte{C_Q} \atransition{g\recep y} Q' \spcontexte{C_Q'}}
            {P \para Q \spcontexte{C} \tautransition P' \para Q' \spcontexte{C'}}
$$
where
\begin{itemize}
\item $x \in \varpile s \cup \varpile{s_P}$, $x\in q$
\item $y \in \varpile s \cup \varpile{s_Q}$, $y\not\in q$, $y$ of type $\qubittype$
\item$C' = \contexte{\rmpile{(\pileajoutp s {(s_P\para s_Q)})}{\ensemble x}}
{q[x\leftarrow y]}{\rho}{f}$
\end{itemize}

$$
\frac{P \spcontexte{C_P} \dtransition P' \spcontexte{C_P'}
             \hspace{20pt}
              Q \spcontexte{C_Q} \dtransition Q' \spcontexte{C_Q'}}
            {P \para Q \spcontexte{C} \dtransition \stopproc \spcontexte{C'}}
$$
with
$C' = \contexte{s}{\codomsoustq{q}{e}}{\traceout{e} q \rho}{\domsoust{f}{e}}$ and
$e = (\varpile{s_P}\cup\varpile{s_Q})$

\subsubsection*{Variable declaration}
$$
\frac{}{\debutbloc \declvar{x_1}{t_1},\ldots,\declvar{x_1}{t_n} \findeclvar  P \finbloc \spcontexte{C}
\transition \debutbloc P \finbloc \spcontexte{C'}}
$$
with $C = \contexteqcq$, $C' = \contexte{s'}{q}{\rho}{f}$\\
and $s' = \pileajoutp s {\ensemble{(x_1,t_1),\ldots,(x_n,t_n)}}$

\subsubsection*{End of scope of variables}
$$
\frac{P \spcontexte{C} \divtransition P' \spcontexte{C'}}
{\debutbloc P \finbloc \spcontexte{C} \divtransition \debutbloc P'\finbloc \spcontexte{C'}}
$$
where $\divtransitionse$ stands for any of the transitions: $\petiteatransition\alpha$
with $\alpha$ an action, $\petitetautransition$, or $\transitionse$.
$$
\frac{P \spcontexte{C} \dtransition P' \scontexte{\pileajoutp s e}{q}{\rho}{f}}
{\debutbloc P \finbloc \spcontexte{C} \dtransition \stopproc\ 
\scontexte{s}{\codomsoustq{q}{e}}{\traceout{e}{q}{\rho}}{\domsoust{f}{e}}}
$$

\end{document}